\documentclass[a4paper,prl,preprint,nofootinbib]{revtex4}
\usepackage[]{graphicx}
\begin{document}

\title{Histogram Reweighting Method for Dynamic Properties}

\author{Carlos Nieto-Draghi\footnote{Present address: Institut Fran\c{c}ais du P\'etrole, 1-4 Avenue de Bois Pr\'eau, 92852 Rueil-Malmaison, France}, Javier P\'erez-Pellitero, and Josep Bonet Avalos\footnote{Corresponding author}}
\affiliation{Departament d'Enginyeria Qu\'{\i}mica, ETSEQ, Universitat
  Rovira i Virgili, Avda. dels Pa\"{\i}sos Catalans 26, 43007 Tarragona Spain}

\date{\today}
\begin{abstract}
  The histogram reweighting technique, widely used to analyze Monte Carlo data, is shown to be applicable to dynamic properties obtained from Molecular Dynamics simulations. The theory presented here is based on the fact that the correlation functions in systems in thermodynamic equilibrium are averages over initial conditions of functions of the trajectory of the system in phase-space, the latter depending on the volume, the total number of particles and the classical Hamiltonian. Thus, the well-known histogram reweighting method can almost straightforwardly be applied to reconstruct the probability distribution of initial states at different thermodynamic conditions, without extra computational effort. Correlation functions and transport coefficients are obtained with this method from few simulation data sets.
\end{abstract}

\pacs{05.10.-a,02.70.Ns,66.20.+d}
\keywords{histogram reweighting, molecular dynamics, transport coefficients, correlation functions}

\maketitle
 
The histogram reweighting (HR) technique~\cite{Ferrenberg1988,Ferrenberg1989} is, at present times, widely used in the determination of properties of thermodynamic equilibrium of many-body systems from a reduced amount of Monte Carlo data. It is recognized that the use of HR techniques increases the accuracy in the evaluation of thermodynamic properties and reduces the computer time needed for precise estimates of these properties. In this letter we address the problem of how the HR method can be extended to equilibrium dynamic properties, like correlation functions and transport coefficients, from molecular dynamics simulations.

In a many-particle system in contact with a reservoir keeping constant a given set of intensive thermodynamic variables, the fluctuations in the conjugate extensive variables permit the system to explore thermodynamic conditions that differ from those fixed by the reservoir. The {\em a priori} knowledge of the functional form for the probability distribution of these fluctuations then permit to reconstruct, {\em reweight}, the histogram of the density of states of the system from a collection of Monte Carlo data sets. Then, the desired thermodynamic averages in different thermodynamic conditions can be evaluated. In the following, we will discuss how the same principle applies for dynamic properties of systems in thermodynamic equilibrium and show some results of the application of the method.

	The dynamics of a classical N-body physical system in phase-space is described by the Liouville equation~\cite{Goldstein80}
\begin{eqnarray}
\frac{\partial}{\partial t}P(\Gamma,t)&=&\sum_{i=1}^{N} \left(\frac{\partial H}{\partial \vec{r}_i}\cdot \frac{\partial }{\partial \vec{p}_i}-\frac{\partial H}{\partial \vec{p}_i} \cdot \frac{\partial }{\partial \vec{r}_i}	\right)P(\Gamma,t)\nonumber \\
& & \equiv -i{\cal L}P(\Gamma,t) 		\label{1}
\end{eqnarray}
where $P(\Gamma,t)$ is the full phase-space probability distribution, $H(\{\vec{p}_i\},\{\vec{r}_i\})$ is the Hamiltonian of the system, while $\vec{r}_i$ and $\vec{p}_i$ are, respectively, the position and momentum of the i$^{th}$-particle. We are implicitly assuming that there are no external time-dependent force fields and that positions and momenta are generalized coordinates, for simplicity. We denote by $\Gamma$ a given point in phase-space, i.e., to the ensemble of positions and momenta of all the particles, $\{\vec{p}_i\},\{\vec{r}_i\}$. ${\cal L}(\Gamma)$ is the Liouville operator which governs the evolution of the probability distribution in phase-space. If we denote by $A(\Gamma)$ and $B(\Gamma)$ given phase-space functions, we can evaluate the correlation function from the two-event probability distribution, which can be recast in terms of the single-event probability and the conditional probability of two events, to yield
\begin{equation}
\langle A(t)B(0)\rangle = \int d\Gamma \int d\Gamma' A(\Gamma)B(\Gamma') P(\Gamma,t|\Gamma',0) P(\Gamma',0)
		\label{2}
\end{equation}
In this expression, $P(\Gamma',0)$ is the probability distribution of the system at the initial time, while the conditional probability $P(\Gamma,t|\Gamma',0)$ stands for the probability of finding the system in the state $\Gamma$ at the time $t$, given that it was in the state $\Gamma'$ at $t=0$. If the system is in thermodynamic equilibrium then $P(\Gamma',0)=P(\Gamma',t)=P_{eq}(\Gamma')$. The form of the latter is known {\em a priori} and is characterized by the set of intensive variables kept fixed by the reservoir, which determine the actual thermodynamic state. For simplicity, we will only deal with the canonical ensemble, thus $P_{eq}(\Gamma')= \exp(-H[\Gamma']/kT)/Z$, with $Z=\int d\Gamma' \exp(-H[\Gamma']/kT)$, $k$ being the Boltzmann's constant. In eq. (\ref{2}) $P_{eq}(\Gamma')$ is in fact the distribution of initial states of the subsequent dynamics of the system. It is crucial in our analysis to realize that $P_{eq}(\Gamma')$ contains all the required thermodynamic information of the system in the calculation of correlation functions. Effectively, the conditional probability  $P(\Gamma,t|\Gamma',0)=\exp(-it{\cal L})\delta(\Gamma-\Gamma')$ depends only on phase space variables. This conditional probability can in addition be written in terms of a trajectory in phase-space $\Gamma[t;\Gamma']$, i.e. $P(\Gamma,t|\Gamma',0)=\delta(\Gamma-\Gamma[t;\Gamma'])$, due to the fact that the Liouville operator simply introduces a displacement in the $\Gamma$ coordinates, according to the Hamilton's equations of motion for a deterministic conservative dynamics. In these expressions, $\Gamma[t;\Gamma']$ is the set of positions and momenta of all the particles at the time $t$, which depend on the actual time as well as on the initial positions and momenta, $\Gamma'$. Hence, after integration over $\Gamma$ the correlation function takes the form
\begin{equation}
\langle A(t)B(0)\rangle = \int d\Gamma' \, A(\Gamma[t,\Gamma'])B(\Gamma') P_{eq}(\Gamma')
		\label{3}
\end{equation}
Eq. (\ref{3}) shows that $\langle A(t)B(0)\rangle$ depends on an {\em equilibrium} average of the function of the trajectory $\Psi[t;\Gamma'] \equiv A(\Gamma[t,\Gamma'])B(\Gamma')$ over the initial conditions. The thermodynamic information is therefore contained only in the specification of these initial conditions. The function $\Psi[t;\Gamma']$ progresses along a trajectory according to the action of the Hamiltonian $H[\Gamma]$. If the Hamiltonian is time-independent, then the energy is therefore a constant of motion during the trajectory and so they are $V$ and $N$. As a consequence, the evaluation of correlation functions corresponds to an ensemble average of trajectories characterized by constant values of $N$, $V$, and $E$, weighted with the probability distribution of the initial states corresponding to the actual thermodynamic ensemble. 

Therefore, from the simulation point of view, let us assume that we perform a set of different NVE molecular dynamics runs from specified initial conditions, randomly selected from a given thermodynamic ensemble. The values of the function $\Psi[t;\Gamma']$ are recorded, together with the values of the extensive parameters that characterize the statistical weight of the initial conditions $\Gamma'$ of the given trajectory in the chosen ensemble. In the particular case of the canonical ensemble, the thermodynamic state is then characterized by $N,V$ and $T$. Thus, we will record $\Psi[t;\Gamma']$, together with $E=H[\Gamma']$ for every sample, due to the fact that $P_{eq}(\Gamma') \propto \exp (-H[\Gamma']/kT )$, in this case. In fact, an usual constant temperature MD simulation can be performed by using standard methods. The thermostat (a weak coupling with a bath) can be switched on, to appropriately set the initial state according to the suited canonical distribution, and off, during the measurement of the trajectory function $\Psi[t;\Gamma']$. Then, in a single simulation, the correlation function is obtained as usual from the simulated data, i.e.
\begin{equation}
\langle A(t)B(0)\rangle = \int d\Gamma' \, P_{eq}(\Gamma') \Psi[t;\Gamma'] \simeq \frac{1}{\Lambda}\sum_{i=1}^{\Lambda} \Psi[t;\Gamma_i']	\label{4}
\end{equation}
In this expression, $\Gamma_i'$ is the i$^{th}$ sample obtained in the simulation and $\Lambda$ is the total number of samples. It is then clear that $\Psi[t;\Gamma']$ is a phase-space function that depends on one additional parameter, i.e. the time. Therefore, $\Psi$ is suitable for the application of the HR technique provided that the sampling of the initial state distributions are propertly done. 

Effectively, in the standard application of the HR method, one would aim at reconstructing the order parameter probability distribution
\begin{equation}
P(\Psi(t),T)=\int d\Gamma' \, P_{eq}(\Gamma', T) \delta(\Psi(t)-\Psi[t;\Gamma'])	\label{5}
\end{equation}
at given thermodynamic conditions, from MD estimates of $\Psi$. Since we are discussing the Canonical ensemble case, we have explicitly indicated the temperature $T$ in eq. (\ref{5}) and in what follows, for clarity reasons. Hence, from a set $\alpha=1 \dots N$ of independent molecular dynamics simulations, characterized each one by its temperature, $T_\alpha$, following the standard application of the HR method, one would write~\cite{Frenkel} 
\begin{equation}
P(\Psi(t),T)d\Psi(t)=\frac{\sum_{\alpha=1}^{N}dn_\alpha(t)P(\Psi(t),T)}{\sum_{\alpha=1}^{N}\Lambda_{\alpha}P_{\alpha}(\Psi(t),T_\alpha)} \label{6}
\end{equation}
where $\Lambda_{\alpha}$ is the total number of samples in the $\alpha^{th}$ simulation and $dn_{\alpha}(t)$ stands for the number of samples with a value of the variable in the range $(\Psi(t),\Psi(t)+d\Psi(t))$ in the $\alpha^{th}$ simulation. Moreover, $P_{\alpha}(\Psi(t),T_\alpha)$ is the order-parameter probability distribution in the $\alpha^{th}$ data set. Finally, the desired correlation function is obtained from the reconstructed probability
\begin{eqnarray}
\langle A(t)B(0)\rangle &=& \int d\Psi(t) \, \Psi(t) P(\Psi(t),T) \simeq \nonumber \\
 & \simeq & \sum_{\Psi(t)} \sum_{\alpha=1}^{N} \left( \frac{ dn_\alpha(t) \Psi(t) P(\Psi(t),T)}{\sum_{\alpha=1}^{N}\Lambda_{\alpha}P_{\alpha}(\Psi(t),T_\alpha)} \right)
		\label{7}
\end{eqnarray}
Eq. (\ref{7}), nevertheless, poses the difficulty of constructing the histograms for the variable $\Psi(t)$. However, one can formally avoid the computation of histograms by inverting the order of summation to recover the sum over states\cite{Newman}. One then finally writes
\begin{equation}
\langle A(t)B(0)\rangle \simeq  \sum_{\alpha=1}^{N} \sum_{\Gamma_{i_\alpha } '} \left( \frac{ \Psi[t;\Gamma_{i_\alpha }'] P(\Gamma_{i_\alpha }',T)}{\sum_{\alpha=1}^{N} \Lambda_{\alpha}P_{\alpha}(\Gamma_{i_\alpha } ',T_\alpha)}\right)
		\label{8}
\end{equation}
where $P_\alpha(\Gamma')= \exp(-H[\Gamma']/kT_\alpha)/Z_\alpha$, with $Z_\alpha=\int d\Gamma' \exp(-H[\Gamma']/kT_\alpha)$ and similarly with $P(\Gamma ',T)$, for the example discussed in this letter. As in the usual application of the HR method, the set of partition functions $Z_\alpha$ has to be obtained following the standard procedure, which customary implies iteratively solving the corresponding integrals~\cite{Frenkel,Newman}. Eq. (\ref{8}) is the most important result of this letter. On one hand, permits to obtain the value of a correlation function at a temperature $T$ from a set of simulations made at temperatures $T_\alpha$. The expression is not limited to the canonical ensemble, but other ensembles can be analogously treated. On the other hand, the use of generalized Einstein relations or Green-Kubo expressions permit the evaluation of transport coefficients from the correlation functions.

\begin{figure}[ht]
\vspace{0.5cm}
\includegraphics[width=6cm,height=6cm]{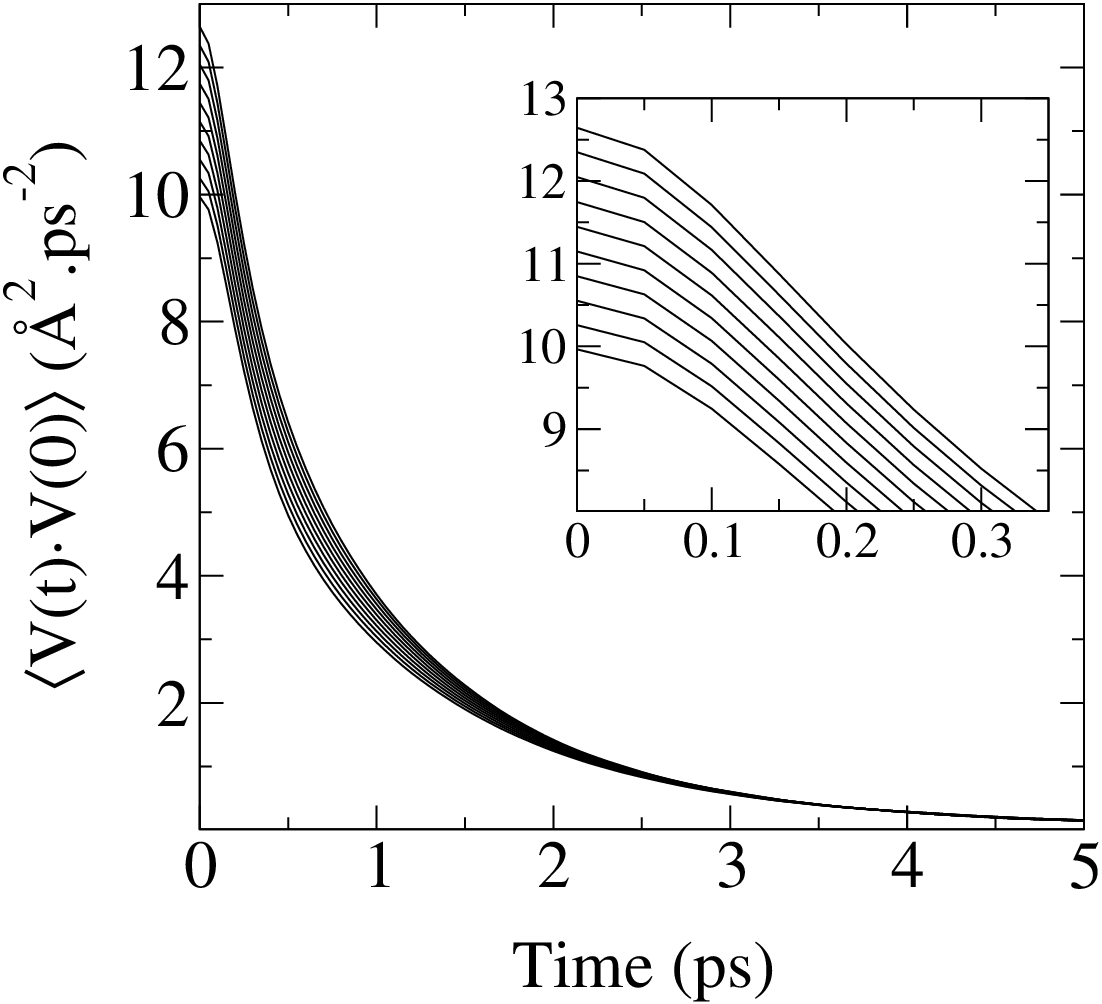}
\caption{\label{VACF}\small Velocity autocorrelation function (VACF) for $10$ temperatures in the range $158.157K-202.657K$, selected every $5K$, as obtained by means of the HR technique. The inner plot shows the initial behavior of the VACF.  \normalsize}
\end{figure}

\begin{figure}[ht]
\includegraphics[width=6cm,height=6cm]{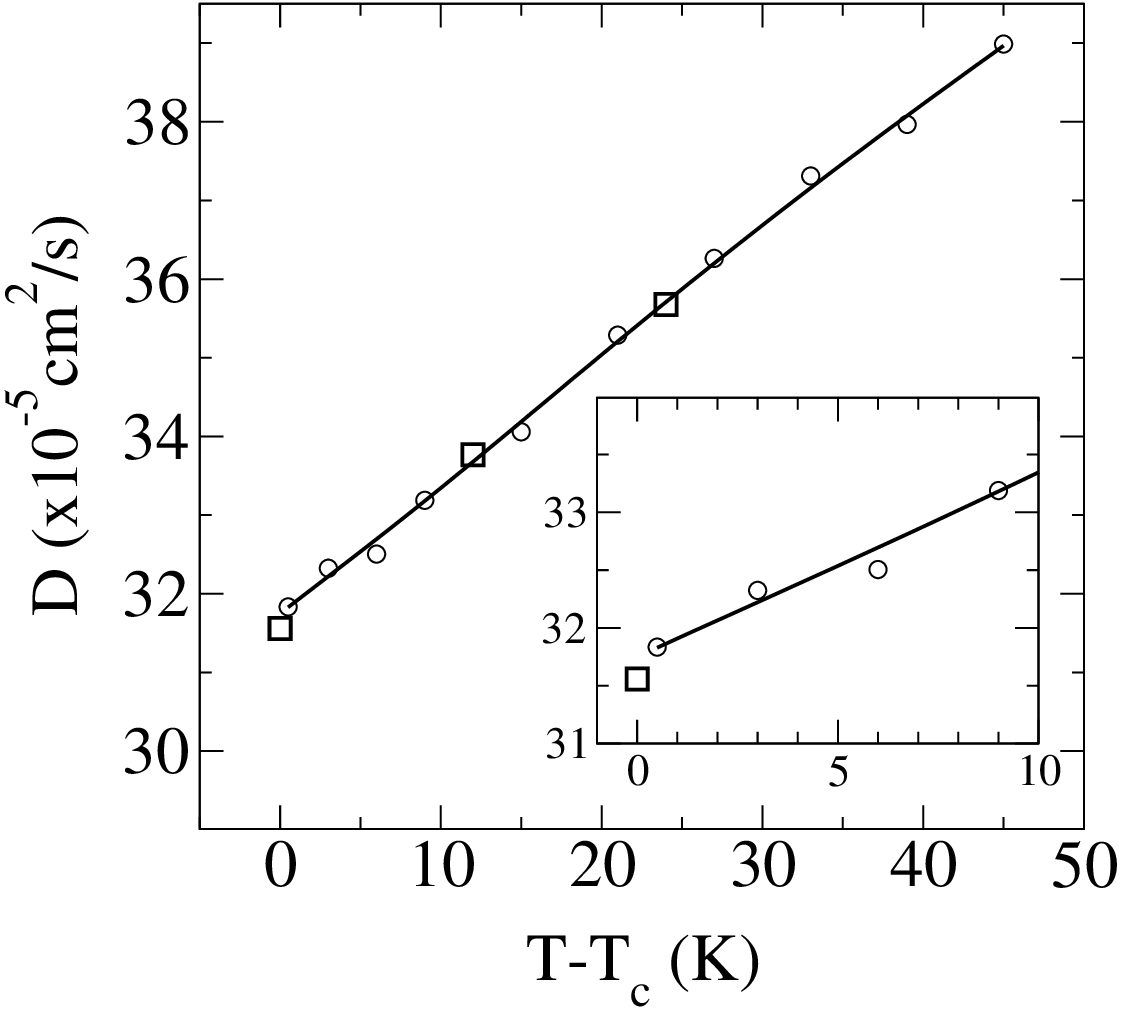}
\caption{\label{D} \small Self-diffusion coefficient. The circles are the values of $D$ obtained from the $10$ MD data sets through the Green-Kubo formalism. The solid line are the estimates obtained from the application of the HR method using these data sets. Squares are independent MD simulations at $T=T_c, \; 169.657 K$ and $181.657 K$, not included in the HR. The incept shows the behavior of $D$ near the critical point.\normalsize}
\end{figure}

To illustrate the theory developed here, we present results for some correlation functions and transport coefficients of supercritical Argon near the critical point. The simulated system consists of $N=256$ Lennard-Jones particles with parameters $\varepsilon = 0.9961 KJ/mol $ and $\sigma= 3.405 \AA$, in a cubic box of size $L=32.15646 \AA$ so that the density of the system is $\rho_{c}=510.7136 kg/m^3$, which corresponds to the critical isochore of the model used, being the critical temperature $T_c=157.657 K$~\cite{Li}. We have performed $10$ MD simulations at temperatures ranging from $T_1=158.157K$, slightly above the critical point, up to $T_2=202.657K$. 
The simulations are performed thermostating the system via a Nos\'e-Hoover procedure\cite{Frenkel} during $5 ps.$, corresponding to $5000$ time-steps. After this period, the energy of the initial state is recorded and a constant $N,V,E$ integration is carried out for additional $15 ps.$, to obtain a sample of the suited function $\Psi[t;\Gamma']$ along the simulated trajectory. The process is repeated up to $2500$ samples are obtained per simulation. Then the $10$ MD data sets are used in eq. (\ref{8}) to interpolate the values of the desired correlation function at $50$ different evenly distributed temperatures in the temperature range between $T_1$ and $T_2$. In Fig.\ref{VACF} we can observe the  velocity autocorrelation function for some selected temperatures in the analyzed domain. Upon integration of the velocity autocorrelation function, the corresponding Green-Kubo formula yields the diffusion coefficient of the system~\cite{Hansen}, obtained as a continuous function of the temperature. The results are given in Fig. \ref{D}.


\begin{figure}[ht]
\vspace{0.5cm}
\includegraphics[width=6cm,height=6cm]{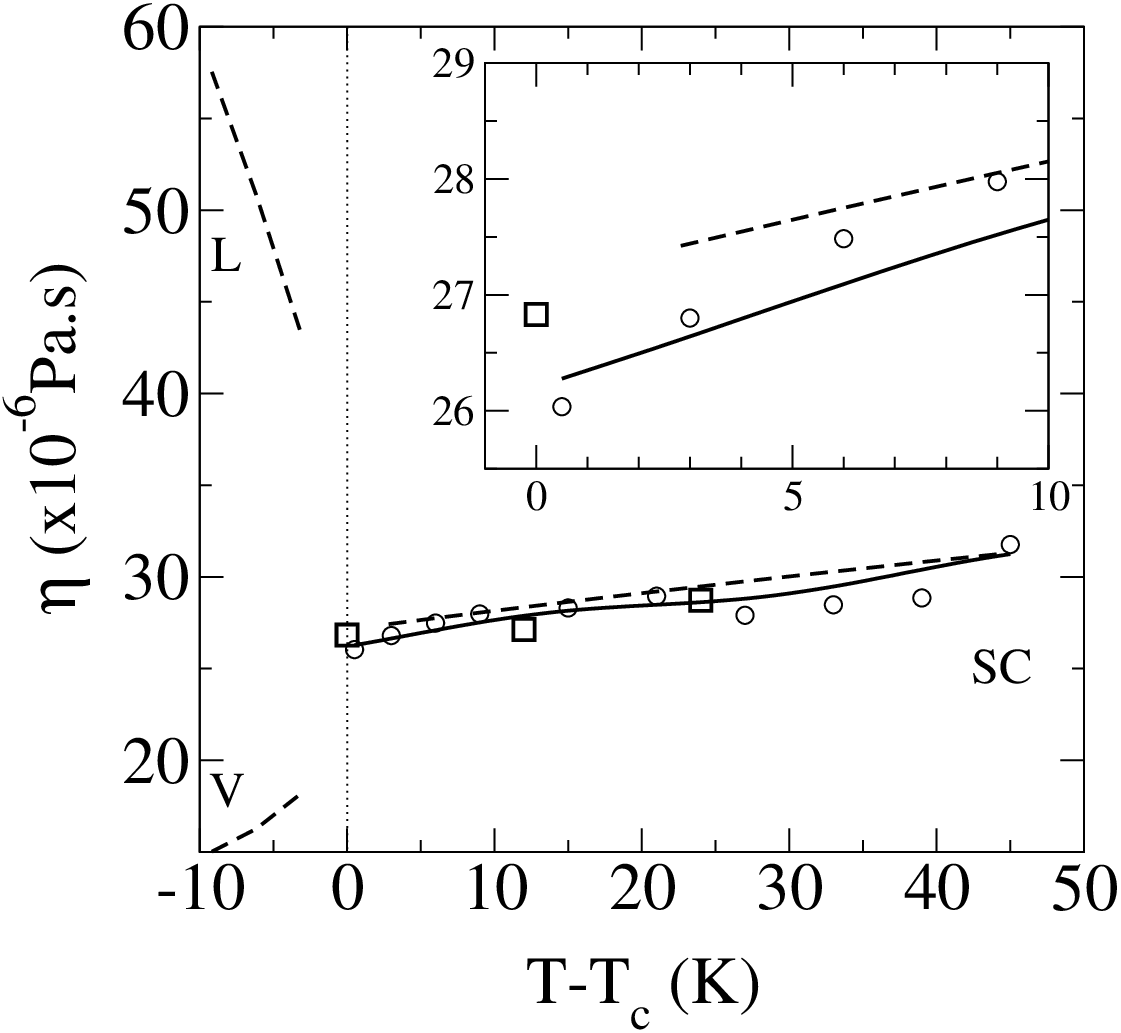}
\caption{\label{etaArgon}\small Viscosity coefficient for Argon at supercritical conditions. The simbols have the same meaning as in Fig. \ref{D}. Dashed line corresponds to the experimental values\cite{Exp}. The inner plot shows the behavior of the shear viscosity close to the critical point. \normalsize}
\end{figure}

The evaluation of the viscosity of the system as a continuous function of the temperature has been carried out through an Einstein relation that involves the time-integral of the stress tensor\cite{Dysthe}
\begin{equation}
\eta = \frac{1}{20} \frac{V}{k_{B}T} \lim_{t\to\infty}  \frac{d}{dt}
\left[ \sum_{\alpha}\left<\Delta P_{\alpha\alpha}\left(t\right)\right>^{2}
+2\sum_{\alpha > \beta} \left< \Delta P_{\alpha\beta} \left(t\right) \right>^{2}\right] \label{eta}
\end{equation}
Here, $\alpha$ and $\beta$ are indexes running over the three Cartesian coordinates, $V$ is the volume, and $\Delta P_{\alpha\beta} (t)$ denotes the {\it displacement} of the elements of the pressure tensor $P_{\alpha \beta}$
\begin{equation}
\Delta P_{\alpha\beta}\left(t\right) = \! \! \int^t_0 \! d\tau \! \frac{1}{2}\!  \left(\! P_{\alpha\beta}\left(\tau\right)+
P_{\beta\alpha}\left(\tau\right)- \frac{2}{3} \delta_{\alpha \beta}
\sum_{\beta}  P_{\beta\beta}\left(\tau\right)\! \right) \label{P}
\end{equation}
%
%

\noindent where

\begin{equation}
P_{\alpha\beta}\left(t\right) = \frac{1}{V} \left(\sum_{i} \frac{p_{\alpha i}(t)
p_{\beta i}(t)}{m_{i}} + \sum_{i<j} \sum f_{\alpha ij}(t) r_{\beta ij}(t)\right) \label{P_micro}
\end{equation}

\noindent is the microscopic expression of the stress tensor\cite{Hansen}. In eq. (\ref{P_micro}), $p_{\alpha i}$ is the $\alpha$-component of the momentum of particle {\it i}, $f_{\alpha ij}$ is the $\alpha$-component of the force exerted on particle {\it i} by particle {\it j}, and $r_{\beta ij}$ is the $\beta$-component of the particle-particle vector, ${\bf r}_{ij} \equiv {\bf r}_j-{\bf r}_i$. Here, the trajectory function is $\Delta P_{\alpha\beta}\left(t\right)$, according to eq. (\ref{P}). After the proper averages at a desired temperature are made using eq. (\ref{8}), then eq. (\ref{eta}) yields the corresponding value of the viscosity coefficient at this temperature. Repeating this procedure for each temperature, one obtains a continuous curve, as can be seen in Fig. \ref{etaArgon}. 
The agreement between the simulated and empirical data\cite{Exp} is very good. In the strict vicinity of the critical point the results are obviously affected by finite-size effects that impeach the weak divergence of this coefficient at the critical point to be reproduced by our simple simulations\cite{Jutta}. For the same reason, no anomaly in the self-diffusion coefficient is observed\cite{Drozdov}. A deep analysis of the dynamic behavior of the system at the critical point is beyond the scope of this letter. However, we want to stress that the method developed here might be very useful in the analysis of the dynamic critical behavior when applied in combination with finite-size scaling procedures.

Finally, we want to point out that the HR method implemented here is sensitive to the quality of the energy fluctuations produced by the thermostating method. In particular, we have observed that the energy fluctuations obtained from the method of velocity rescaling of Berendsen\cite{Berendsen} significantly differ from the Boltzmann distribution that we have assumed in the application of eq. (\ref{8}). As a consequence, the predictions near and out of the edges of the temperature range simulated, almost entirely constructed on these fluctuations, yield rather poor estimates of the physical quantities evaluated.

Therefore, data on dynamic properties from MD simulations can be combined to extract dynamic information on the system at different thermodynamic conditions. We have shown that the method is suitable for the evaluation of correlation functions as well as the determination of transport coefficients through both Green-Kubo and Einstein relations. In particular, the agreement between the experimental values of the viscosity coefficient and the HR predictions from simulations is very good. In fact, as expected, the HR data shows a smoother behavior than the single simulation values of this coefficient since a larger amount of data is used in the interpolation than in the single simulation evaluation. The method is not limited to the Canonical ensemble but any other ensemble can be used if the proper probability distribution for the fluctuating extensive variable is {\em a priori} known. 

The authors are indebted to Dr. Allan D. Mackie for many fruitful discussions. This work has been supported by the Spanish Ministerio de Educaci\'on y Ciencia, grant No. CTQ2004-03346/PPQ.

\end{document}